\documentclass[letterpaper,twocolumn,10pt]{article}

\pdfoutput=1

\usepackage{fancyhdr}
\usepackage{hyperref}
\usepackage[width=7in,left=.75in,height=9.5in,top=.75in]{geometry}
\usepackage{graphicx}
\graphicspath{{figures/}}
\usepackage{listings}
\usepackage{color, soul}
\usepackage{comment}
\usepackage{multirow}
\usepackage{url}

\definecolor{codegreen}{rgb}{0,0.6,0}
\definecolor{codered}{rgb}{0.7,0,0}
\definecolor{codegray}{rgb}{0.5,0.5,0.5}
\definecolor{codepurple}{rgb}{0.58,0,0.82}
\definecolor{codeblue}{rgb}{0,0,0.8}
\definecolor{backcolour}{rgb}{0.95,0.95,0.98}
\lstdefinestyle{mystyle}{
  backgroundcolor=\color{backcolour},   commentstyle=\color{codered},
  keywordstyle=\color{codegreen},
  numberstyle=\tiny\color{codegray},
  stringstyle=\color{codeblue},
  basicstyle={\footnotesize\ttfamily},
  language=C,
  breakatwhitespace=false,         
  breaklines=true,                 
  captionpos=b,                    
  keepspaces=true,                 
  numbers=left,                    
  numbersep=5pt,                  
  showspaces=false,                
  showstringspaces=false,
  showtabs=false,                  
  tabsize=2
}

\usepackage[normalem]{ulem}

\newcommand{\ignore}[1]{}
\newcommand{\function}[1]{\texttt{#1}}
\newcommand{\ResetModel}{Reset Model}
\newcommand{\Manual}{Manual Model}
\newcommand{\Auto}{Automatic Model}
\newcommand{\NameShort}{PDRM}
\newcommand{\FullName}{Persistent Data Retention Model}
\newcommand{\OriginalProgram}{\textsc{P-Original}}
\newcommand{\RetentionProgram}{\textsc{P-Retain}}
\newcommand{\ManualProgram}{\textsc{P-Manual}}
\newcommand{\AutoProgram}{\textsc{P-Auto}}

\newcommand{\LEDS}{LEDS}
\newcommand{\LEDSLong}{Lazily Extendable Data Structures}

\begin{document}

\title{\FullName{}s
}

\author{\normalsize Tiancong Wang and James Tuck \\ \normalsize NC State University \\ \normalsize \{twang14,jtuck\}@ncsu.edu}

\date{}
\maketitle

\thispagestyle{empty}

\begin{abstract}

Non-Volatile Memory devices may soon be a part of main memory, and programming models that give programmers direct access to persistent memory through loads and stores are sought to maximize the performance benefits of these new devices. Direct access introduces new challenges. In this work, we identify an important aspect of programming for persistent memory: {\em the persistent data retention model.}

A Persistent Data Retention Model describes what happens to persistent data when code that uses it is modified. We identify two models present in prior work but not described as such, the Reset and Manual Model, and we propose a new one called the Automatic Model. The Reset model discards all persistent data when a program changes leading to performance overheads and write amplification. In contrast, if data is to be retained, the Manual Model relies on the programmer to implement code that upgrades data from one version of the program to the next. This reduces overheads but places a larger burden on the programmer. 

We propose the Automatic Model to assist a programmer by automating some or all of the conversion. We describe one such automatic approach, Lazily Extendible Data Structures, that uses language extensions and compiler support to reduce the effort and complexity associated with updating persistent data. We evaluate our \NameShort{}s in the context of the Persistent Memory Development Kit (PMDK) using kernels and the TPC-C application. \Manual{} shows an overhead of 2.90\% to 4.10\% on average, and \LEDS{} shows overhead of 0.45\% to 10.27\% on average, depending on the workload. \LEDS{} reduces the number of writes by 26.36\% compared to \Manual{}. Furthermore, \LEDS{} significantly reduces the programming complexity by relying on the compiler to migrate persistent data. 

\end{abstract}

\section{Introduction}
\label{section:introduction}

Recently, Non-Volatile Memory (NVM) technologies are garnering attention because they provide persistent storage, byte-addressability, and have reasonably fast access latency~\cite{3dxpoint,miccheck,pcm,spram,sttram,reram}. These NVM technologies may be adopted as part of the main memory hierarchy, and adoption of non-volatile main memory (NVMM) may have a large impact on the way future computer systems are designed and programmed.

In conventional systems, persistent storage is accessed through the file system. In contrast, NVMM offers the potential for programmers to access persistent data directly through loads and stores. Providing {\em direct access} (DAX) to programmers improves performance, by cutting away unnecessary layers of software, but also introduces challenges, like support for failure-safety~\cite{persistency,bpfs,pbarrier,pmemio} and the need for position independence~\cite{chen-micro50}.

In the pursuit of direct access, many have proposed that persistent data will be created and manipulated directly by programs using conventional declarations, data structures, and pointers~\cite{nvheaps,mnemosyne,nvmdirect,wang-micro50}. Data can be placed in persistent memory on one run of a program and then accessed again on a subsequent run. However, this creates a semantic challenge for programmers with regard to how persistent data is described and manipulated over time. Since persistent data is retained across runs of a program, does a declaration of a type refer to the persistent data created in a previous run of the program or to the organization of the data on the next run of the program? In systems with a conventional memory and storage hierarchy, programmers trust that data types describe what will happen when the program runs next, because there is no data already in memory -- memory is repopulated from scratch each run. However, in the context of persistent memory the declarations describe both -- what's already in persistent memory and what will happen on the next run.

Consider an analogy with files on conventional systems without NVMM. If a file format changes, the software is modified to support reading the old file type and converting it as needed into a new format. For data stored serially on secondary storage, this is reasonable since it must be moved from disk to memory on each invocation of the program. But, persistent memory is different in that persistent structures never need to be serialized. However, if the type declarations used in the previous execution change, then the address calculations for the persistent data accessed in the next execution will be invalid or erroneous. Current languages and compilers have no automatic way of detecting the discrepancy or helping the programmer deal with it.\footnote{There are lot's of software-based solutions here. The programmer could add fields to the data structures, like a magic word, to track versions and manually convert between types on any change to the code. However, this is an onerous and error-prone requirement to put on all persistent data.} 

The responsibility of understanding what's already in persistent memory and how to handle changes falls squarely to the programmer. Few works~\cite{nvmdirect,pmemio}, if any, have considered this problem. In PMDK and other frameworks, the programmer will keep the old declaration, add a new declaration to support the needed changes, and then copy from the old data structure into the new one on a subsequent run of the program. This approach leads to redundantly declared structures and harmful write amplification due to copied data. 

To begin addressing this problem, we propose that programming environments define a {\em \FullName{}} (\NameShort{}) that explains how persistent data is {\em retained} as programs are modified and change over time. Programmers need, at a minimum, a clearly specified model of behavior.

Perhaps the simplest model would be one that discards all persistent data any time code changes. In a sense, this is equivalent to treating the persistent data the same way we think of DRAM being cleared on each run of the program, except that the data in NVM would be cleared on each modification and re-compilation of the program.
We call it the {\em \ResetModel{}}. While simple, it implies that programmers cannot rely on data remaining in persistent memory after making changes to their code, which defeats the purpose of persistent data structures and is inconsistent with the goal of supporting direct access to NVMM. It may also causes write amplification, since data will be recalculated and stored into memory again.

Alternatively, we can make the programmer fully responsible for distinguishing such data and managing it across changes to the code. When a program is changed in a way that affects persistent data structures, the programmer should manually convert the data that is affected by the layout change into a new format, so we call it the {\em \Manual{}}. While programmers take full responsibility in this model, some minimal supports are needed from the environment to ensure that it is supported. This is supported in the PMDK because all persistent objects are contained in pools, but this is not supported for static persistent variables in Mnemosyn because there is no way to refer to the value held in a static persistent variable before it was modified.   

Last, we define the {\em \Auto{}}. The \Auto{} assists programmers in the conversion of data and ensures the data continues to be used appropriately after code changes. This requires a mechanism that can distinguish which data can be retained (and how to) and which cannot after a program is modified. We describe one technique, Lazy Extendible Data Structures (LEDS), that fits into this classification that uses language and compiler support to assist the programmer in automatically updating the layout of data structure, as we expect this to be a common need. Furthermore, for the use cases it supports, we show that it is competitive with an optimized {\em \Manual{}} based approach while requiring less programmer effort.

Other useful \NameShort{}s may exist, in particular hybrid models that are composed of these three, but we do not consider them in depth in this work. 

We make the following contributions: 1) We identify the problem of persistent data retention and describe three different \FullName{}s, the \ResetModel{}, the \Manual{} and the \Auto{} represented by \LEDS{}. The \ResetModel{} and \Manual{} are present, to some degree in prior work, but \LEDS{} is new to this work. 2) We examine the implementations of the \NameShort{}s in the context of the Persistent Memory Development Kit (PMDK, formerly known as NVML)~\cite{pmemio}, and we implemented and evaluated \LEDS{}. 3) We evaluate the \Manual{} and \LEDS{} with kernels written for PMDK and the TPC-C application where the \Manual{} shows an overhead of 2.9\% to 4.10\% on average and \LEDS{} shows and overhead of 0.45\% to 10.27\%, on average, depending on the workloads. We also find that the number of writes is reduced by 26.36\%, on average, when comparing \LEDS{} to \Manual{}. For TPC-C, \LEDS{} reduces the number of writes by 7.2$\times$ over \Manual{}.

The rest of the paper is organized as follows. Section~\ref{section:motivation} gives a comparison of the traditional file-based persistent programming with persistent memory programming models for NVMM and also explores the \NameShort{}s in existing persistent programming extensions. Section~\ref{section:design} introduces the definition of \ResetModel{}, \Manual{} and \Auto{} in general persistent programming languages and points out key differences between \Manual{} and \Auto{}.  Section~\ref{section:implementation} demonstrates implementations of each model built on PMDK libraries. Section~\ref{section:methodology} presents the environment setup for the experiments and explains the workloads and some terminologies used in experiments. Section~\ref{section:evaluation} demonstrates the overhead of \Manual{} and \Auto{} and analyzes the components of the overhead and also characterizes the two models. Section~\ref{section:relatedwork} presents related work and Section~\ref{section:conclusion} concludes the paper.

\section{Motivation}
\label{section:motivation}

\subsection{Comparison between persistent programming and file-based programming}



Let's compare traditional file-based support for persistent data and new persistent memory programming models. As shown in Figure~\ref{fig:compare-persistent-programming}(a), with traditional file-based programming models, there are two copies of the data existing in the program: one in volatile memory for use in computation and one in a file. The interpretation of the two copies is written by the programmer. Here, the struct clearly only describes the copy in memory, not storage.

For persistent memory, as shown in Figure~\ref{fig:compare-persistent-programming}(b), the programmer only needs to define a data structure and denote that it resides in NVMM. The struct itself now describes storage\footnote{Here we assume some persistent programming model like Mnemosyne. Other persistent programming models like NV-Heaps and PMDK can also achieve this with more complicated interfaces.}. 
Since we are considering persistent programming models that provide direct access to programmers, there's only one copy of the data structure and it's persistent.

\begin{figure}[h]
\begin{lstlisting}[language=C, title={(a) Traditional file-based programming},captionpos=b, numbers=none]
struct example{
    int x;
    int y;
};
struct example array[100];
int main(){
    // Interpreter
    FILE* fp = open("somefile.txt", "r");
    while (fscanf(fp, "%d %d", 
                  &array[i].x, &array[i].y) != EOF){
        i++;
    }
    fclose(fp);
    // Some operations
    ...
    // Another interpreter to save data to files
}
\end{lstlisting}
\begin{lstlisting}[language=C, title={(b) Persistent programming},captionpos=b, numbers=none]
struct example{
    int x;
    int y;
};
persistent struct example array[100];
int main(){
//Some operations on array
...
}
\end{lstlisting}
\caption{A comparison of having a persistent data structure using (a) traditional file-based programming models and (b) using persistent programming models. \label{fig:compare-persistent-programming}}
\end{figure}

However, it also introduces a problem when a program changes the struct definition. In the programs in Figure~\ref{fig:compare-persistent-programming}, suppose the programmer wants to add a new field \texttt{int z} to the \texttt{struct example}. It's easy for traditional models in (a). They can modify the volatile data structure freely because the struct does not describe the file contents. However, with persistent programming model in (b), the persistent data structure is laid out according to the original struct (two integers), but after the change the program believes the persistent struct has three integers. Thus it will read the wrong data if it continues using the old data still held in persistent memory.

There are many reasonable software approaches to fix this problem.
We may want to reset the persistent memory (\ResetModel{}). However, the two integer fields \texttt{x} and \texttt{y} of the struct might still be useful and should not be discarded.
For example, a programmer might only add a field in a well-established data structure. Thus all the other fields in the data structure should still be useful.
Furthermore, if discarded, they may simply be recomputed leading to detrimental write amplification and lowering the lifetime of the device. Instead, the \Manual{} and the \Auto{} allow the programmer to retain the persistent data but with different complexity and overhead.

Hence, the key issue is this: declarations now may describe persistent data. As a result, modifications to declarations for persistent data need additional attention by the programmer to ensure that 
the persistent data can evolve as expected when program version evolves.
We want to minimize programmer effort while retaining high performance and write endurance.


\subsection{\NameShort{}s in existing persistent programming models}
\label{section:motivation-existing}

We found no literature that discusses a similar concept as \FullName{}s. However, there is recognition that such support is needed.
Table~\ref{tab:existing-models} summarizes the support in existing persistent programming models. All the libraries we examined could support \ResetModel{} because they use a file abstraction for persistent data.

For NV-Heaps and PMDK, \Manual{} can be implemented by manually duplicating the data structure and copying from an old struct layout to a new one. NVM-direct \cite{nvmdirect}
identified the problem of needing to upgrade data structures and even suggests
allocating the struct with more memory than needed to support in-place updates.
They also define an \texttt{upgrade()} function that can perform an in-place upgrade on the struct. However, the function may fail if there's not enough space for the upgrade and will force a reset.  

\begin{table}[]
\resizebox{\columnwidth}{!}{
\begin{tabular}{|p{7em}|p{5em}|p{5em}|p{5em}|}\hline
\bf{Programming Models} & \bf{\ResetModel{}} & \bf{\Manual{}} & \bf{\Auto{}} \\ \hline\hline
Mnemosyne~\cite{mnemosyne} & Yes & Partial; Only for dynamic allocations & No \\ \hline
NV-Heaps~\cite{nvheaps} & Yes & Yes & No \\ \hline
NVM-direct~\cite{nvmdirect} & Yes & Yes; \texttt{upgrade()} function & No \\ \hline
PMDK~\cite{pmemio} & Yes; Also, can detect a change of struct layout and resets & Yes & No \\ \hline
\end{tabular}
}
\caption{\NameShort{}s in existing persistent programming models that supports user-defined data types.}
\label{tab:existing-models}
\end{table}

\subsection{PMDK Library}
Our implementations are based on Persistent Memory Development Kit (PMDK, formerly known as NVML)~\cite{pmemio}.  It provides a collection of libraries for various use cases, which is tuned and validated to production quality and thoroughly documented~\cite{pmemio}. It targets server-class applications that requires skilled programmers to write programs with low-level interfaces. There are 10 libraries in the PMDK and we will use the interfaces exclusively from libpmemobj as described in the man page~\cite{libpmemobj}. 
The main idea of PMDK is to store persistent regions in file abstractions, or {\em pools}. Pools are directly mapped to a program's address space and programmers can refer to persistent data with persistent pointers called ObjectIDs, or PMEMoid, which are the equivalent of addresses to persistent memory. 

In PMDK, all persistent data is held within pools. There are no persistent global variables, as in Mnemosyne~\cite{mnemosyne}, that are outside of a pool. Each pool has a root object, which is defined as a struct, to store variables associated with the pool, some of which may be pointers to data structures in the pool. To modify persistent data, a coder would modify one or more struct definitions contained within the persistent pool. Therefore, in our \Auto{} approach (Section~\ref{section:implementation}), we will focus only on dealing with struct layout changes.

PMDK offers some features to help detect if a pool will be misinterpreted. Each pool can be tagged with a layout name when it's created. If the pool is opened with a layout name that's different than the one used in creation, the library will generate an error and abort the program. 
\section{\FullName{}s}
\label{section:design}

As discussed in Section~\ref{section:motivation}, persistent data is tied to their program in persistent programming. In this section, we describe three different models to retain persistent data when a program evolves from one version to another. We only briefly describe \ResetModel{} and \Manual{} because they reflect the state-of-the-art.

\subsection{\ResetModel{}}
{\em \ResetModel{}} mandates that the persistent data region is cleared/reset for a modified program in order to prevent data misinterpretation. The \ResetModel{} is compelling for its simplicity, effectiveness, and soundness. It's detrimental due to write amplification from recomputing data and rebuilding data structures. Write amplification is most severe when the working set of a program is much larger than the amount of data affected by the data structure change.

It's easy for programming models to support \ResetModel{} as long as they can easily clear the persistent regions. For example, for all the persistent programming languages in Table~\ref{tab:existing-models}, they use files to hold persistent data. So when a program is modified, the file should be deleted so the persistent data of the program is cleared. Note, this does not happen automatically, even though it is a necessity for some of the systems to function properly.

\subsection{\Manual{}}

The {\em \Manual{}} requires programmers to transform all the related persistent data to reflect the changes in a newer version of program. 

A sufficient requirement to provide the \Manual{} is that persistent data is dynamically allocated in a persistent heap. Then programmers can allocate new persistent objects, copy the old ones, and fix-up all of the relevant pointers between persistent objects. The \Manual{} may not work for persistent data that is statically mapped by the compiler, like in Mnemosyne, unless an interface is provided for re-linking the data to a new location, which has been proposed in other contexts~\cite{hicks-dsu} and could be applied here.

The \Manual{} can be implemented as a separate conversion program to be executed during re-compilation or as part of the larger program, running only when data layout changes occur. The potential advantage of \Manual{} is that programmers have the freedom to either retain or reset data precisely as needed. However, this approach places a large burden on the programmer. The burden of \Manual{} makes sense if persistent data is either rarely used or predominantly read-only.  

\subsection{\Auto{}}

We wish to make retention of data simpler than in the \Manual{}, thus we propose {\em \Auto{}}. Ideally, \Auto{} would fully or partially automate the update process from one version of code to the next, requiring no special intervention. We believe a wide variety of techniques may fall into this model. On a path toward that goal, we propose Lazily Extendable Data Structures (\LEDS{}).

\subsection{Lazily Extendable Data Structures}

Using new language support, we only require that programmers provide a high level description of how each structure is extended from its previous definition. The extensions allow the structure to grow to encompass new fields or pointers. The compiler analyzes the description and inserts code that will upgrade each object from the old definition to the new one when it is encountered during a subsequent program execution. 

An example of possible C/C++ language extensions are shown in Figure~\ref{fig:migrate-auto-def}. The \texttt{extendible} keyword identifies the struct as one that may be extended in the future. The \texttt{EXTENSION} block
can be added once to indicate new fields that are desired. Within the block, any number of new fields may be added, including a nested \texttt{EXTENSION} block that holds another set of extensions. The \texttt{INIT} block defines how to initialize the new fields from existing fields. 

Each \texttt{EXTENSION} block represents one change or \textit{upgrade} to the data structure. In our current design, extension blocks can only be added and should never be removed or changed, as this would again lead to a mismatch between the struct and the persistent data in memory.

\begin{figure}[h]
\begin{lstlisting}[language=C, 
title={First version of struct A.}, numbers=none]
extendible struct A{
    int val;
};
\end{lstlisting}
\begin{lstlisting}[language=C, 
title={Second version of struct A.}, numbers=none]
extendible struct A{
    int val;
    EXTENSION{
        double val_dbl;
        INIT(A* obj){
            // initialize new field using old val
            val_dbl = obj->val;
        }
    }
};
\end{lstlisting}
\caption{An example of retention with \Auto{}: struct A adds a floating-point type field to the struct A in the second version, and copy the integer value. \label{fig:migrate-auto-def}}
\end{figure}

Based on the extension blocks, the compiler inserts checks into the code to test if an upgrade is needed or not at each access to an object of the same type. How this check is done will be clear after we explain the upgrade process.

If an upgrade is needed, we do not actually allocate a new object and copy over the old one, as in the \Manual{}. Instead, for all structs marked with the \texttt{extendible} keyword, we embed an extra unused pointer field at the end of the struct. If a programmer adds an extension, we allocate it as a new object in the same persistent pool and store its address in the pointer. This means we do not make a copy, and hopefully do not significantly increase the number of writes.  It also means we do not need to fix-up references to the upgraded object held elsewhere in memory.
However, accesses to the fields in the extension have to use an additional load through the pointer to access the desired field, which adds overhead.

The presence of this pointer field makes it simple to test if an object has already been upgraded.  If the pointer is NULL, it has not yet been updated; otherwise, it has been. 

Our approach supports arbitrarily deep nesting. Any new extension field is always given an extra pointer field to point to the next extension.

\begin{figure}[t]
\centering
\includegraphics[width=\columnwidth]{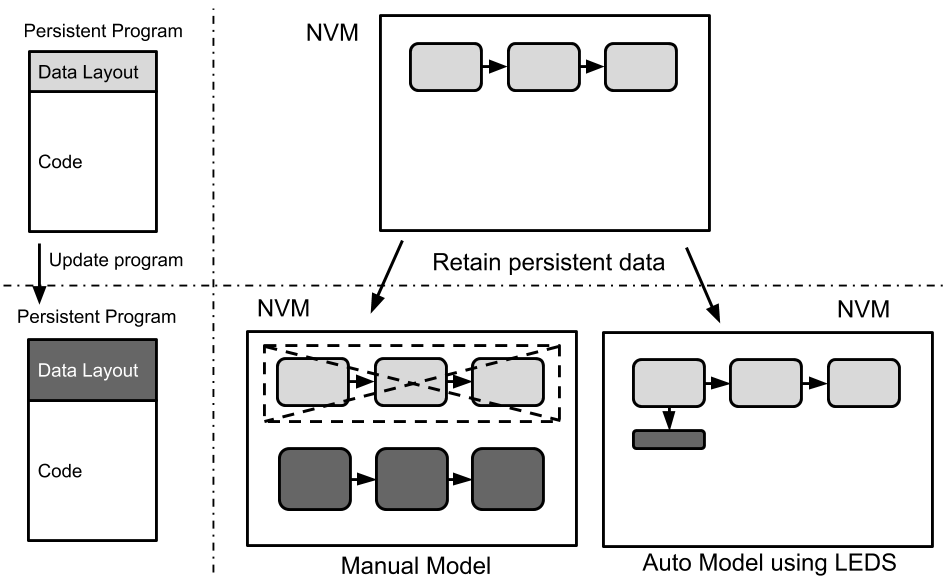} 
\begin{minipage}{\columnwidth}
\caption{A description of the differences in \Manual{} and \Auto{} using LEDS when retaining persistent data in a changed program. The example shows a linked list with three nodes and searching in the updated linked list hits the first node. In \Manual{}, a new linked list needs to be created. And in \Auto{} using LEDS, only the node accessed in the new program (first node) is being updated.}
\label{fig:pdrms}
\end{minipage}
\end{figure}

A comparison between \Manual{} and \Auto{} is shown in Figure~\ref{fig:pdrms}. Note that \Manual{}, in general, would make a copy of the existing data structure.  However, in \Auto{}, we extend the existing data by allocating a linked node.

\paragraph{What makes LEDS lazy?}
Another advantage of our design is that we do not need to track down all linked objects and convert them all at once, as a manual technique might choose to do. Instead, we only convert the objects 
when their changed fields happen to be
accessed during one run of the program. 
Because the language contains the information for how to convert the object, we can convert lazily as we encounter objects with NULL extension fields. Even if multiple nested extensions are needed, they can simply be chained in sequence.

\section{Implementations on PMDK}
\label{section:implementation}

In this section, we discuss the implementations of three \NameShort{}s in libpmemobj in PMDK~\cite{pmemio}. 

\subsection{\ResetModel{}}
To support the \ResetModel{}, we manually delete all pools used by a program before the next execution.

\subsection{Implementation of \Manual{}}

PMDK supports the \Manual{}. To convert persistent data, the coder must read out the persistent data using the current layout from the pool, transform it into a new layout by copying it into a new pool or a different object in the same pool, just like what we would expect from the file-based approach. Figure~\ref{fig:pmdk-manual} gives an example of adding field to a persistent linked list using the libpmemobj~\cite{libpmemobj} in PMDK.


\begin{figure}[]
\begin{lstlisting}[language=C, numbers=none]
struct old_LL {
    int val;
    TOID(old_LL) next; // TOID: PMEMoid with a specified type, used with D_RO and D_RW
};
struct LL {
    int val;
    float val_fl; // Added field in the new version.
    TOID(LL) next;
};
struct old_root_t{
TOID(old_LL) ll_head;
};
struct new_root_t{
TOID(LL) ll_head;
};
// Retain the linked list
TOID(LL) linkedlist_retain(PMEMobjpool* pop, const TOID(old_LL)& old_head)
{
    TOID(LL) head = TX_ZNEW(LL); // TX_ZNEW: Allocate a new object inside a transaction
    D_RW(head)->val = D_RO(old_head)->val; // D_RO and D_RW: translates a PMEMoid type into an address, and reads/write from it
    D_RW(head)->val_fl = D_RO(old_head)->val; 
    TOID(old_LL) old_p = D_RO(old_head)->next;
    TOID(LL) p = head; // Retain the next node of p
    TX_FREE(old_head); // TX_FREE: Free an object inside a transaction
    while(!TOID_IS_NULL(old_p)){
        D_RW(p)->next = TX_ZNEW(LL);
        // Retain the content from old_p to D_RW(p)->next
        D_RW(D_RW(p)->next)->val = D_RO(old_p)->val;
        D_RW(D_RW(p)->next)->val_fl = D_RO(old_p)->val;
        // Move on to the next node in the linked list and free the old node
        TOID(old_LL) to_free = old_p;
        old_p = D_RO(old_p)->next;
        p = D_RO(p)->next;
        TX_FREE(to_free);
    }
    return head;
}
// Retain the root object
TOID(new_root_t) root_retain(PMEMobjpool* pop)
{
    TOID(old_root_t) old_root = POBJ_ROOT(pop, old_root_t); // POBJ_ROOT: Obtain the root object of a pool
    // Create a temporary holder of the new root
    TOID(new_root_t) new_root_temp = TX_ZNEW(new_root_t);
    // Copy the whole linkedlist pointed by the head field
    D_RW(new_root_temp)->head = linkedlist_retain(pop, D_RO(old_root)->head);
    // Allocate a new root and copy from the temp
    TX_FREE(old_root);
    TOID(new_root_t) new_root = POBJ_ROOT(pop, new_root_t);
    TX_MEMCPY(new_root, new_root_temp); // TX_MEMCPY: Perform memcpy()-like operations from one PMEMoid to another inside a transaction
    return new_root;
}
\end{lstlisting}
\caption{An example manual retention of a linked-list when adding a floating point field and retaining the value from the integer field, with native PMDK libraries. \label{fig:pmdk-manual}}
\end{figure}

This example reveals that 
retaining only one field in the root object, which is a linked data structure, takes a non-trivial amount of effort from a programmer, and it has proven to be error prone in the context of PMDK. We make some observations from this example. First, some of this code is general enough that it could be put into libraries. For example, the function \function{root\_retain} is for creating a new root object and retaining fields from an old one. Most of the code applies generally to other programs: it needs to allocate a new root object, read out all the fields from the old root and copy to it. Since there's only one root object in a pool, we should use a temporary holder to hold the contents of a new root before actually making it the official root object. The only part that might be different is how to retain each field (e.g. \function{linkedlist\_retain}). Thus if a function is created for each field that needs retention, \function{root\_retain} can automatically call those functions on each field. The other observation is that the compiler/libraries can't fully automate this whole process. It's because \function{linkedlist\_retain} needs to iterate through every \texttt{old\_LL} object in the old linked list in order to create a new linked list. The iteration might be different depending on the data structure (e.g. graph versus linked list) and hard to automate. 
Furthermore, it's also complicated to retain data when there are nested structs. Even if only one struct changes its layout, all the objects related to it should change (e.g. in the example, only \texttt{struct LL} changes but it also leads to convert the root object, which contains a field related to \texttt{struct LL}).
Thus in \Manual{}, we need the programmer to manually iterate through the old data structure but there could be some library and compiler support to generate some common code to reduce the amount of programming. We do not explore this possibility further.


\subsection{Implementation of \LEDS{} }


We also implement \LEDS{} for the PMDK libraries. We have not yet implemented a front-end for the language extensions, because we wanted to justify that such an effort would be worthwhile. Hence, we focus our implementation on accurately measuring the performance and write amplification of the proposed language extensions. For our implementation, we manually implemented the code using the extensions we described earlier and manually lower them into C/C++ language implementations. 


\subsubsection{Lowering the \texttt{extendible} Struct.}

\begin{figure}[h]
\begin{lstlisting}[language=C,
title={(a) Written by programmers.}, numbers=none]
extendible struct LL{
    int val;
    TOID(struct LL) next;
    EXTENSION{
        float val_fl;
        INIT(LL* obj){
            val_fl = obj->val;
        }
    }
};
\end{lstlisting}
\begin{lstlisting}[language=C,
title={(b) Generated by the compiler.}, numbers=none]
struct LL{
    int val;
    PMEMoid to_extend;
};
struct LL_EXT{
    double val;
    PMEMoid to_extend;
    LL_EXT(const LL* obj) {
        val_fl = obj->val;
        to_extend = OID_NULL;
    }
};
\end{lstlisting}

\caption{An example of retention with \Auto{} using LEDS: adding a floating point value and initializing its value from the integer field of last run. \label{fig:migrate-auto-def-lower}}
\end{figure}

An \texttt{extendible} struct is supported by splitting its definition into multiple structs, one for the original struct and one for each \texttt{EXTENSION} block. A pointer is placed at the end of each struct marked with the \texttt{extendible} keyword to point to the next extension. By convention, we refer to this pointer as \texttt{to\_extend}, and it will be used to access the extension indirectly. We are essentially building a linked list from the extension fields.

Extensions are enforced by the compiler to be strictly nested, to reflect the sequential order of updates to a program. As shown in Figure~\ref{fig:migrate-auto-def-lower}, the original struct LL is split into two structs, LL and LL\_EXT. 


\subsubsection{Lazy updates.} 
The compiler must generate code that upgrades an object on the first access to one of its extended fields. Note, we could attempt to upgrade on any access, but we make this as lazy as possible to reduce the overhead of benign updates to code, for example, adding a field that is not used.

For each struct marked \texttt{extendible}, we traverse the full code and find all references to its fields that are in an \texttt{EXTENSION} block. Just before the access (either a load or store), we insert code to check if the corresponding \texttt{to\_extend} field used to access it is non-NULL. If non-NULL, the object has already been extended and the code falls-through to the access. For he other case, we add code to allocate the extension and insert the initialization code specified in the \texttt{INIT} initializer. Last,
the appropriate indirect access using the \texttt{to\_extend} pointer is generated. 

Since \LEDS{} is a lazy approach, a persistent object might not be accessed after an update is made to its structure definition. In this case, when the programmer accesses fields in the newest extension, the routine also needs to check all the intermediate \texttt{to\_extend} fields that lead to the newest extension in sequence, allocating extensions that turn out to be NULL. For any given extension, the extension only needs to be allocated once, whether in the current run or a later run of the program.


The NULL pointer check on \texttt{to\_extend} is only needed on the first access of one of the fields it directly points to. However, in general, deciding if it is the first access is difficult. Therefore, by default, we always insert a check before any dereference of an extension field. 

Figure~\ref{fig:migrate-auto-def-lower}(a) shows the same data structure from Figure~\ref{fig:pmdk-manual} rewritten for PMDK using \LEDS{}. Now, \texttt{linkedlist\_retain} and \texttt{root\_retain} are no longer needed.

\subsection{Discussion}
\label{section:implementation-auto-discussion}
\subsubsection{Overheads and Optimizations}

\LEDS{} has the potential to incur higher overheads than a manual approach. However, based on our observations, many of these overheads can be reduced through optimization.

\paragraph{Extra Checks.} Because \LEDS{} lazily updates structures, every run of the program will need to check if extendible objects have been updated before accessing any of their extended fields. Since the compiler may not be able to determine the first access of each object, many redundant checks may be inserted. Furthermore, for nested \texttt{EXTENSION} blocks, there will be multiple levels of check if the field is placed in an innermost \texttt{EXTENSION} block.

The compiler can reduce redundant checks on extended objects. For example, a redundant check B can be removed if a check A of the same object is placed in a basic block that dominates B's basic block. We do not evaluate this optimization further in this paper.


\paragraph{ObjectID Translation.}
In PMDK, persistent objects are accessed using ObjectIDs. ObjectIDs must be translated to an address before they can be used to access memory. This overhead can be reduced through compiler or hardware support~\cite{chen-micro50,wang-micro50}. 

The compiler can also reduce the cost of translations by caching them and re-using them. 
Thus it saves some redundant translations. We evaluate this optimization in Section~\ref{section:evaluation-breakdown}.

\paragraph{Unnecessary Copies.} Lastly, the compiler can avoid allocating extensions for objects that only exist in volatile memory for a short period of time. The programmer might create a temporary local variable to hold a copy of persistent data. When copying a field to such a local, we may not need to perform a deep copy if the fields are never modified. We can perform a shallow copy instead. Two of our workloads, namely btree and ctree, benefit from this optimization. We evaluate this optimization in Section~\ref{section:evaluation-breakdown}.

\subsubsection{Liabilities.}

\LEDS{} keeps the original data layout by adding new fields as extensions. This breaks an assumption that programmers and compilers often make: all the data of an object is stored contiguously in the address space. For example, copying structs is a simple memcpy operation. However, that will not work for our extendible structs because they will need a deep copy instead. When a compiler detects a copy, it can insert code as necessary to make a deep copy. However, if a programmer manually copies a struct, it could lead to an error, perhaps adding some difficulty for using \LEDS{}. 

The extension field also adds some complexity in terms of freeing the extendible object. When free() is called on the extendible object, it needs an explicit call to free the extensions as well. The compiler can insert a routine to make sure the extensions of the object are all freed, including the extensions of the extensions. However, again, there may be cases this goes undetected due to pointer casting, allowing some extended objects to leak. However, these dangers are not significantly different than those already present in the C language.


\section{Methodology}
\label{section:methodology}
The evaluation is performed on a workstation summarized in Table~\ref{tab:environment}. We implement the \Manual{} and using interfaces in Persistent Memory Development Kit (PMDK)~\cite{pmemio}, formerly known as NVM Library (NVML), developed by Intel. We also implement \LEDSLong{} (\LEDS{}) as an \Auto{} using PMDK libraries. From now on, we will refer to \LEDS{} as \Auto{}. We run the benchmarks on a hard disk instead of real NVM hardware. PMDK supports hard disk with NVM interfaces and simulates the \texttt{clflush} instructions (used for making sure the data is persistent) with an msync() call. In our experiment, we use the configuration of PMEM\_IS\_PMEM\_FORCE=1 provided by PMDK to avoid issuing msync() and unnecessarily slowing down the program.

\begin{table}[h]
\resizebox{\columnwidth}{!}{
\begin{tabular}{|p{7em}|p{17em}|}\hline
Processor & Intel Core 2 Duo CPU E8400 @ 3.00GHz \\ \hline
CPU Cache & L1D: 32KB, L1I: 32KB, L2: 6MB \\ \hline
Memory & 4GB DRAM \\ \hline
Operating System & Linux version 2.6.32 (Red Hat Enterprise Linux Workstation release 6.9) \\ \hline
Hard Disk & 110G \\ \hline
Library & Persistent Memory Development Kit (PMDK) Version 1.3.1 \\ \hline
\end{tabular}
}
\caption{Summary of the environment for experiments}
\label{tab:environment}
\end{table}

\subsection{Workloads}
\label{section:methodology-workloads}
The workloads are shown in Table~\ref{tab:workloads}. We select multiple implementations of maps (list map, hash map and tree map) from examples/libpmemobj in PMDK as our workloads, and we implement a TPC-C application with PMDK interfaces. 

The original programs are modified to use a 32-bit key\footnote{PMDK kernels implements 64-bit key so we change it to 32-bit, in order to retain the keys in the updated program.} and are referred as \OriginalProgram{}. In order to measure the overhead of different retention situations, we make some changes to the layout (\textsc{Layout-x}). In order to perform retention and preserve the data in the \OriginalProgram{}, we write retentions and updated programs for both the \Manual{} and the \Auto{}. For the \Manual{}, we write a separate program to perform the retention (\textsc{P-Retain}), so the updated program \ManualProgram{} only needs to change the definition in the \OriginalProgram{}. With \Auto{}, we integrate the retention in the updated program \AutoProgram{} using our proposed language features and their required transformation.

\begin{table}[h]
\resizebox{\columnwidth}{!}{
\begin{tabular}{|p{7em}|p{17em}|}\hline
{\bf Workloads} & {\bf Descriptions} \\ \hline
skiplist & Skip list data structure with 4 levels. \\ \hline
ctree & Crit-bit Tree \\ \hline
btree & B-Tree of Order of 8 \\ \hline
rbtree & Red Black Tree \\ \hline
hashmap\_tx & Hash map implementations using transaction APIs, with 10 buckets initially \\ \hline
TPC-C & The implementation is based from an implementations that uses volatile memory. We move the B+ Trees (Order of 8) on to the NVM with PMDK interfaces. \\ \hline
\end{tabular}
}
\caption{Description of the workloads from PMDK}
\label{tab:workloads}
\end{table}

In our experiments, we make two different data layout changes: \textsc{Layout-Change} is that we change the 32-bit key to 64-bit key and \textsc{Layout-Add} is that we add a new field to the node, a char array as the name. This measures the impact of different usage scenarios when changing a field: the key field will be used heavily, whereas the newly added field might be used very lightly. We expect the real world usage will be a combination of the two situations.

The experiments measure the overhead of retention in both models: in \Manual{}, the overhead of retention is the time spent on the retention program \RetentionProgram{}. So we calculate the overhead as: 
\begin{equation} \label{eq:manual}
    Overhead_{Manual} = \frac{T_\textnormal{\sc P-Retain}}{T_\textnormal{\sc P-Manual}}\times100\%
\end{equation}
In \Auto{}, the overhead of retention is the additional runtime needed by \Auto{} over \Manual{}. This measure shows what would have been
possible if the new struct were contiguous and had no overhead to create. We calculate the overhead as: 
\begin{equation} \label{eq:auto}
    Overhead_{Auto} = \frac{T_\textnormal{\sc P-Auto} - T_\textnormal{\sc P-Manual}}{T_\textnormal{\sc P-Manual}}\times100\%
\end{equation}
In our experiments, the overhead is calculated by executing the \RetentionProgram{}, \ManualProgram{} and \AutoProgram{} for five times and calculate the average execution time before feeding into the two equations to calculate overheads.

For the PMDK map kernels, we write a main program that performs all three operations (insertion, deletion and search) on the map randomly: we randomly generate a number of integers and search them in the map. If it's found, we remove the key and value pair from the map. Otherwise, we insert a key-value pair for the integer. The sequence of random numbers is provided as an input to ensure the same behavior across all executions. 
In our experiments, we use three different types of inputs to evaluate different behaviors, \textsc{PMDK-INS}, \textsc{PMDK-DEL} and \textsc{PMDK-RAND}, which is named after the behavior in the update program with explanations in Table~\ref{tab:patterns}. These configurations are used to measure the impact of different sequences of operations in the update program, which might result in different overheads for the retention models. 

\begin{table}[h]
\resizebox{\columnwidth}{!}{
\begin{tabular}{|p{5em}|p{4em}|p{18em}|} \hline
{\bf Name} & {\bf Abbr.} & {\bf Descriptions} \\ \hline
Deletions Only & PMDK-DEL & \OriginalProgram{} inserts N unique keys and \ManualProgram{}/\AutoProgram{} deletes the same N unique keys. \\ \hline
Insertions Only & PMDK-INS & \OriginalProgram{} inserts N unique keys and \ManualProgram{}/\AutoProgram{} inserts another N unique keys. \\ \hline
Combination of Insertions and Deletions & PMDK-RAND & \OriginalProgram{} inserts N random keys that may be repeated and \ManualProgram{}/\AutoProgram{} uses the same input. The update program will have a combination of insertions and deletions. \\ \hline
\end{tabular}
}
\caption{Description of different configurations of inputs to the benchmarks in PMDK.}
\label{tab:patterns}
\end{table}

\begin{table}[h]
\resizebox{\columnwidth}{!}{
\begin{tabular}{|p{5em}|p{6em}|p{16em}|} \hline
{\bf Category} & {\bf Term} & {\bf Explanation} \\ \hline
\multirow{2}{5em}{Different layout changes} & \textsc{Layout-Change} & Change the field \texttt{key} from 32-bit to 64-bit. \\ \cline{2-3}
& \textsc{Layout-Add} & Add a new field of a char array. \\ \hline
\multirow{4}{5em}{Different programs in the experiments} & \OriginalProgram{} & The original version of program with 32-bit keys. \\ \cline{2-3}
& \RetentionProgram{} & The program used to perform retention in \Manual{} and executed before the updated program. \\ \cline{2-3}
& \ManualProgram{} & The program with updated layout definition in \Manual{}. \\ \cline{2-3}
& \AutoProgram{} & The program with updated layout definition and primitives for retention in \Auto{}. \\ \hline
\end{tabular}
}
\caption{Terms used in the experiments.}
\label{tab:terminology}
\end{table}
\section{Evaluation}
\label{section:evaluation}

\subsection{Overhead in \Manual{} and \Auto{}}
\label{section:evaluation-overhead}
\begin{figure}[t]
\centering
\includegraphics[width=\columnwidth]{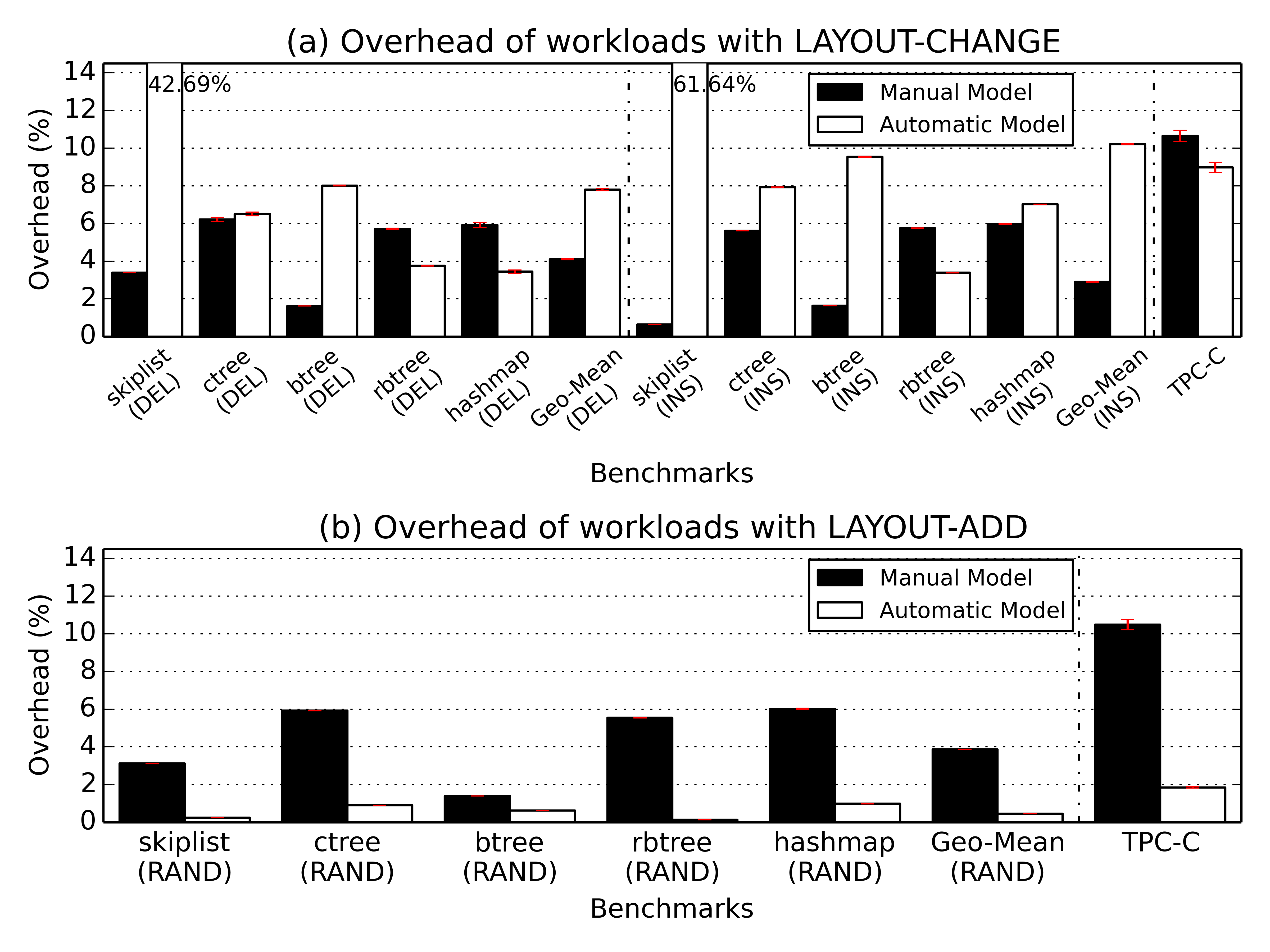} 
\begin{minipage}{\columnwidth}
\caption{Overall performance of the overhead of \Manual{} and \Auto{} on different workloads. (a) \textsc{Layout-Change}: key changes from 32-bit to 64-bit. (b) \textsc{Layout-Add}: add a new field to each node. The result of PMDK kernels is measured with 100000 operations in both \OriginalProgram{} and updated program, and the result of TPC-C is measures with 10 warehouses and 200000 random client operations.}
\label{fig:overheads}
\end{minipage}
\end{figure}

In this section, we present the overhead of the \Manual{} and the \Auto{} with two different \textsc{Layout} changes on the workloads mentioned in Table~\ref{tab:workloads}. The overheads calculated by equations Eq.~\ref{eq:manual} and Eq.\ref{eq:auto} are shown in Figure~\ref{fig:overheads}.

We analyze the PMDK kernels first. For the \Manual{}, both \textsc{Layout-Change} and \textsc{Layout-Add} need to duplicate the old data structure because the size of the node changes. The execution time of \RetentionProgram{} is similar in these benchmarks but the overhead is slightly different due to different execution times for the \ManualProgram{}, as a result of a different number of insertions or deletions. For most workloads, an insertion operation and a deletion operation take a similar amount of time, except for skiplist where skiplist(INS) runs much longer than skiplist(DEL) due to larger data size. Hence, the overhead of retention is amortized in that case. For \textsc{Layout-Change}, \Manual{} has 4.10\% and 2.90\% overhead in PMDK-DEL and PMDK-INS kernels respectively, while \textsc{Layout-Add} has 3.87\% overhead in PMDK-RAND kernels. Note, these overheads are similar.

For \Auto{}, the overhead can vary significantly across workloads. First, \textsc{Layout-Change} has a bigger impact on the overhead (7.81\%, 10.21\% for PMDK-DEL and PMDK-INS) than \textsc{Layout-Add} (0.45\%) because the key is constantly accessed. Hence, it needs checking, redirection, and translation on the field quite often. However, with \textsc{Layout-Add}, it's a newly added field that is not otherwise used, so it adds little overhead. In reality, we expect applications will have a combination of changes akin to \textsc{Layout-Change} and \textsc{Layout-Add} so the performance will be somewhere in-between. 

When comparing the PMDK-DEL and PMDK-INS, \Auto{} can have larger overhead in the latter because there are more keys in PMDK-INS and every access to the key can add overhead. PMDK-INS can also have larger overhead because newly allocated nodes carry extra work to build the extensions. We will break down the overhead of \Auto{} model in Section~\ref{section:evaluation-breakdown}.

When comparing the \Manual{} and the \Auto{} across the workloads, some general trends can be observed. The \Manual{} has similar overhead to retain data whether it's adding new fields or changing existing fields (as long as the size changes) because it needs to allocate a new data structure and copy from the old one. But, the overhead paid in the \Auto{} is relative to the number of accesses of the changed fields. If the accesses are rare, the \Auto{} can have much smaller overhead than \Manual{}. 

As for the TPC-C application, \Manual{} has similar overheads with \textsc{Layout-Change} and \textsc{Layout-Add}, with 10.65\% and 10.49\% overheads respectively. The \Auto{} attains a better performance in these two scenarios, with 8.98\% in \textsc{Layout-Change} and 1.85\% in \textsc{Layout-Add}. 

Furthermore, in TPC-C, the \Auto{} has smaller overhead than the \Manual{}, which shows another advantage of the \Auto{}. The \Auto{} only transforms what is needed instead of transforming all the data. Our experiment shows that the second run of the TPC-C after the update only accesses about 67.4\% of the nodes from the previous run, so \Auto{} only performs 67.4\% of the work that \Manual{} does. We further analyze the sensitivity on the ratio of working data set over total data set in Section~\ref{section:evaluation-sensitivity}. 

\subsection{Breakdown of overhead in \Auto{}}
\label{section:evaluation-breakdown}
\begin{figure}[t]
\centering
\includegraphics[width=\columnwidth]{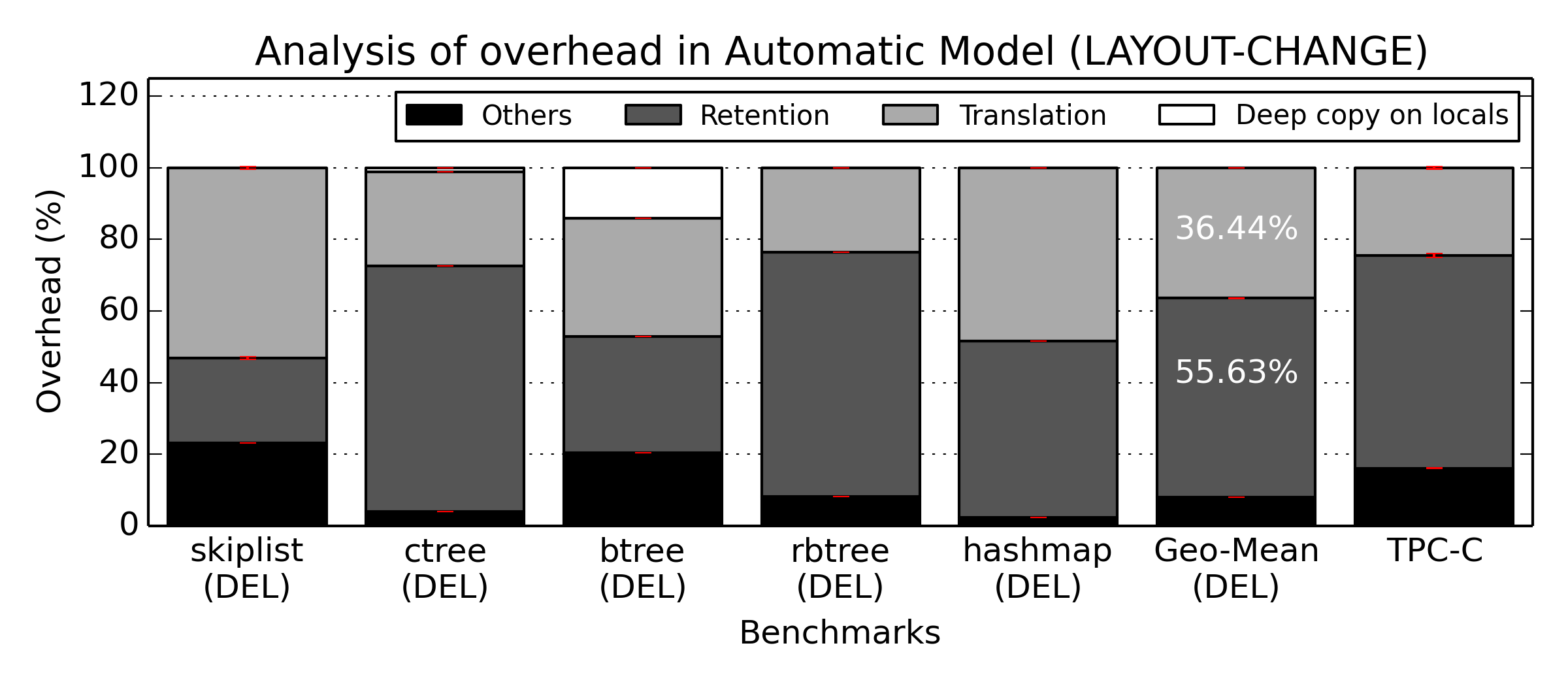} 
\begin{minipage}{\columnwidth}
\caption{A breakdown of the overheads in \Auto{}. The experiment is done with PMDK-DEL workloads of \textsc{Layout-Change}, with 100000 insertions in \OriginalProgram{} and 100000 deletions in \AutoProgram{}.}
\label{fig:breakdown}
\end{minipage}
\end{figure}

There are multiple components of the overheads in \Auto{}, as analyzed in Section~\ref{section:implementation-auto-discussion}. We will evaluate three major sources of overhead: the time for allocating extension object and copying fields, the extra redirection and translation cost to access extensions, and the overhead when copying objects with extensions. To measure the time that \AutoProgram{} programs spent on each overhead, we implement multiple programs that remove each component from the \AutoProgram{} program individually and see how much overhead it can save. Figure~\ref{fig:breakdown} shows the fraction of each component for each workload. 

The most significant component is the cost of allocating the extension and initializing it. 
We see that PMDK-DEL spends on average 55.63\% of the overhead for allocation. 
The second most significant component is the overhead of translation. 
On average PMDK-DEL spends 36.44\% of the overhead for such translations. 
We are using native PMDK translation functions and we think with faster translation techniques like~\cite{wang-micro50,chen-micro50}, this overhead can be reduced. 
It's also worth noting that even with virtual addresses stored in the \texttt{to\_extend} field, accessing a field in extension still has an extra level of redirection because it needs extra loads to locate the extension object first. 
The portion of such overhead is illustrated in the Others part in the Figure~\ref{fig:breakdown} and is non-trivial.

The last component of overhead we analyzed is the cost of making deep copies of objects with extensions. 
Only two workloads ctree and btree have this situation. In btree, 14.10\% of the overhead falls into this category, and we can save the unnecessary allocations in about 51.61\% of the copies.

\subsection{Sensitivity Analysis: Size of data set}
\label{section:evaluation-sensitivity}

\begin{figure}[t]
\centering
\includegraphics[width=\columnwidth]{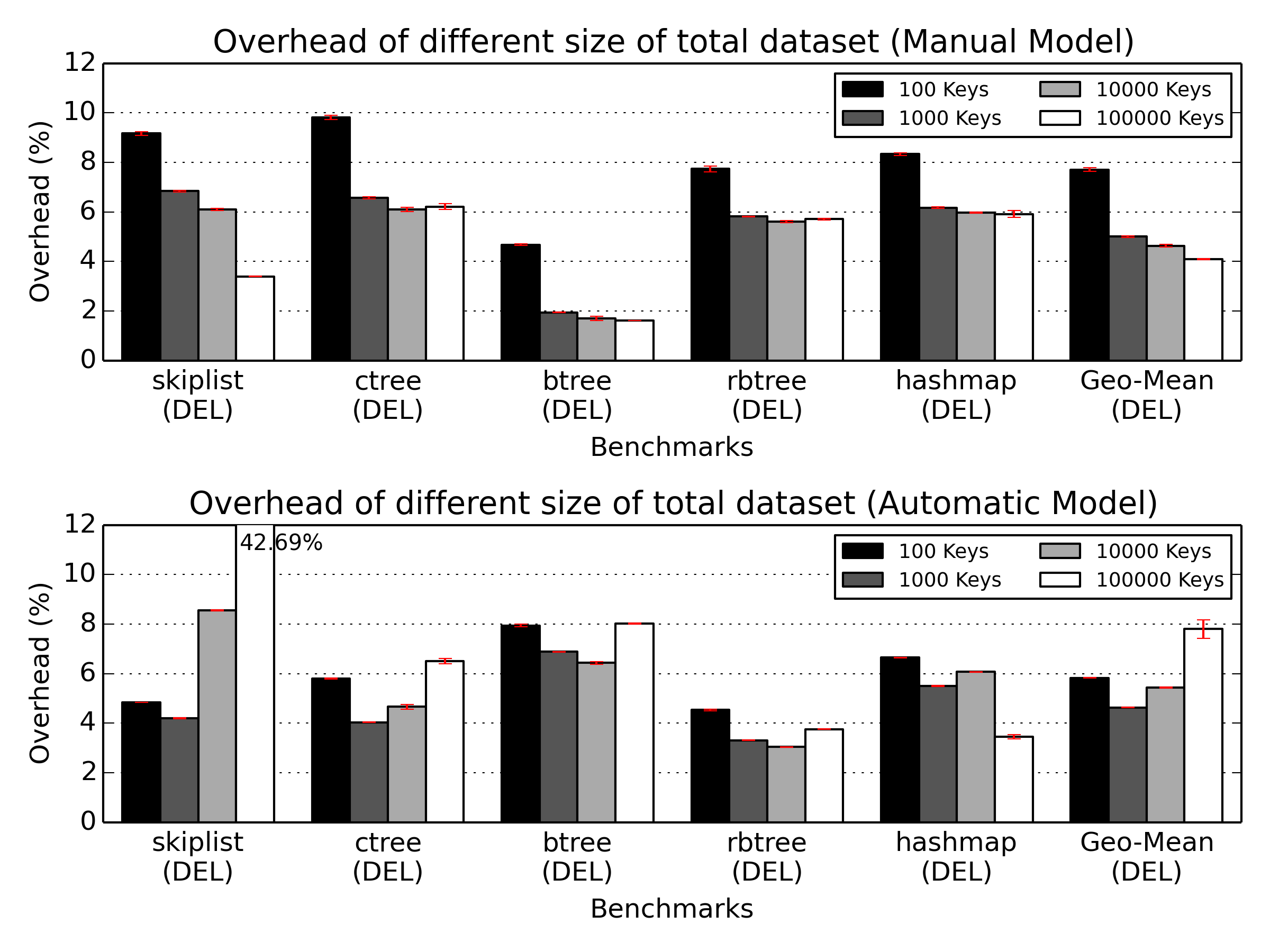} 
\begin{minipage}{\columnwidth}
\caption{Sensitivity analysis of the impact of different data set on both (a) \Manual{} and (b) \Auto{}. The experiment is done with PMDK-DEL of \textsc{Layout-Change}, with different number of operations of 100, 1000, 10000, 100000, resulting in different size of data set for retention. }
\label{fig:manual-size}
\end{minipage}
\end{figure}

Now we consider the impact that the size of the data set has on the overhead. In this experiment, we vary the total amount of data and the total number of operations performed on the data in the same proportion.Figure~\ref{fig:manual-size} shows the overhead of \Manual{} and \Auto{} with multiple data set sizes to retain.

In the case of \Manual{}, for sizes above 100 keys, the behavior is fairly uniform. skiplist is an exception because with increasing \texttt{N} each deletion might take longer to perform due to the larger data structure. This makes the cost of \textsc{P-Retain} relatively less than \OriginalProgram{}.

For the \Auto{}, the results are also fairly uniform, with a slight worsening for large data sizes. The overhead comes from the time spent on allocation and copying new extension objects and also the accesses to the fields placed in extensions. Thus for larger \texttt{N}, each deletion will access a larger number of keys that will increase the overhead. The trend is the most significant for skiplist that needs to access at most \texttt{N} keys for each deletion. Evenso, for some workloads, like rbtree, ctree, and hashmap, \Auto{} performs better than \Manual{}. 

\subsection{Sensitivity Analysis: Ratio of working data set over total data set}

\begin{figure}[t]
\centering
\includegraphics[width=\columnwidth]{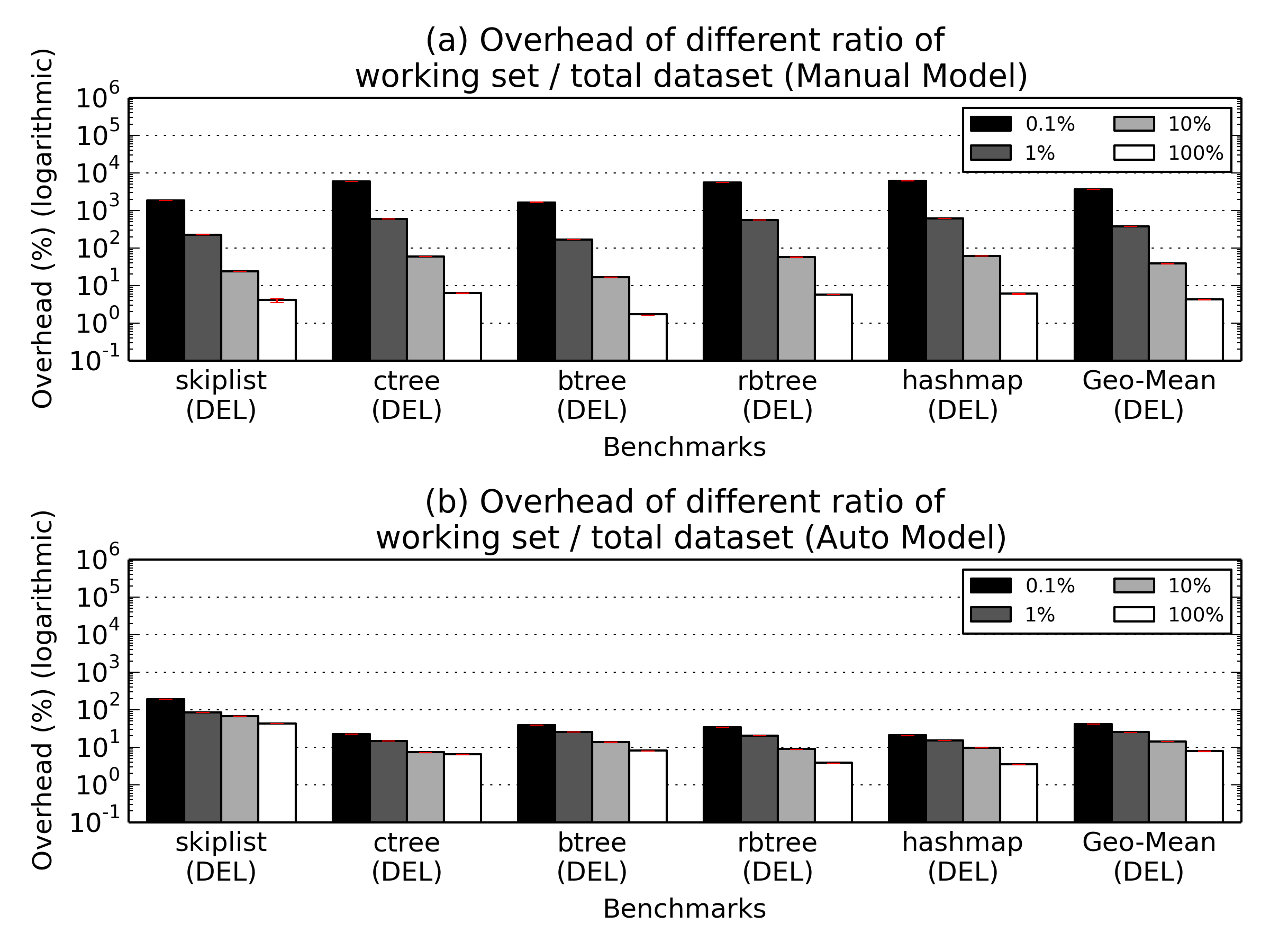} 
\begin{minipage}{\columnwidth}
\caption{Sensitivity analysis of different ratio of working data set in updated program on both (a) \Manual{} and (b) \Auto{}. The experiment is done with PMDK-DEL of \textsc{Layout-Change}, where \OriginalProgram{} performs 100000 insertions and update program deletes 0.1\%, 1\%, 10\% and 100\% of the keys respectively. The y-axis is on logarithmic scale.}
\label{fig:auto-operations}
\end{minipage}
\end{figure}

The \Auto{} only transforms data that is accessed. For programs with a small working set compared to their total data set, \Auto{} has an advantage. We vary the number of deletions performed in the update program \ManualProgram{} and \AutoProgram{} and present the overhead of both \Manual{} and \Auto{} in Figure~\ref{fig:auto-operations}. 
For small working sets, \Auto{} significantly outperforms
\Manual{}, even for the 10\% case. 

\subsection{Write Amplification}

NVMM technologies have limited write endurance, hence techniques that increase writes are undesirable. Reducing the number of writes is always prudent. 

In this section, we estimate the number of bytes written under the \Manual{} and the \Auto{} when converting old data structure. Table~\ref{tab:write-endurance} shows the number of additional writes incurred by the \Manual{} and the \Auto{} for PMDK-DEL and TPC-C on the \textsc{Layout-Change} workload.

\begin{table}[h]
\centering
\resizebox{\columnwidth}{!}{
\begin{tabular}{|p{4em}||p{3em}||p{11em}|p{3em}|p{3em}||p{3em}|p{3em}|} \hline
 & {\bf Reset} & \multicolumn{3}{|c||}{\bf Manual} & \multicolumn{2}{|c|}{\bf Auto} \\ \hline		
 & {\bf Total Bytes (MB)} & {\bf Struct in retention program} & {\bf Each mig. (B)} & {\bf Total Bytes (MB)} & {\bf Each mig. (B)} & {\bf Total Bytes (MB)} \\ \hline
{\bf skiplist} & 9.9 & struct \{{\bf uint64\_t}, PMEMoid\}, TOID{[4]} & 88 & 8.4 & 40 & 3.8 \\ \hline
{\bf ctree} & 6.4 & int, struct \{{\bf uint64\_t}, PMEMoid\} {[2]} & 52 & 4.9 & 40 & 4.3 \\ \hline
{\bf btree} & 7.4 & int, struct \{{\bf uint64\_t}, PMEMoid\} {[8]}, TOID{[8]} & 324 & 6.7 & 40 & 9.5 \\ \hline
{\bf rbtree} & 8.8 & {\bf uint64\_t}, PMEMoid, enum, TOID, TOID {[2]} & 76 & 7.2 & 40 & 3.8 \\ \hline
{\bf hashmap} & 4.6 & {\bf uint64\_t}, PMEMoid, TOID & 40 & 3.8 & 40 & 3.8 \\ \hline \hline
{\bf TPC-C} & 94.2 & unsigned, {\bf uint64\_t} {[8]}, PMEMoid {[8]} & 196 & 81.0 & 40 & 11.1 \\ \hline
\end{tabular}
}
\caption{Number of writes for different \FullName{}s after 100000 insertions in \OriginalProgram{}.}
\label{tab:write-endurance}
\end{table}

For the \ResetModel{}, the data of original program is discarded after 100000 runs. There's retention of the data. When the program changes, we need to repeat 100000 runs and during each run, we need to repeat the writes to NVM as the original program. Table~\ref{tab:write-endurance} shows an estimation of the number of bytes written when the program is re-executed.

In \Manual{}, when we modify the key field from 32-bit to 64-bit, we need to create new objects. Also, this adds extra writes for the fields that do not change since they are copied. Different workloads have different structures to hold the key-value pair.

Table~\ref{tab:write-endurance} shows the definition of each struct that needs retention in the second column. The \{uint64\_t and PMEMoid\} pair is the key-value pair, and note that we need to retain all the structs that have the key-value struct as its field. The size of each struct is calculated in the second column\footnote{The size of PMEMoid and TOID is 128-bit (16 Bytes), and TOID is a typed PMEMoid type in PMDK}. All the bytes that are written during the \RetentionProgram{} are calculated in the third column. 

Meanwhile, with the \Auto{}, we only need to allocate data for the fields placed in the extension, which in this case is the 64-bit key. We also need the \texttt{to\_extend} field in the extension in order for further extension, resulting in 24 bytes total. 
We also need to write the extension field of the object (with the allocated ObjectID), which is 16 Bytes. Thus in \Auto{}, each retention needs to write 40 Bytes of data, and the total number of bytes to write is calculated in the last column. From the results in column 3 and 5, we can conclude that \Auto{} writes fewer bytes to NVM than \Manual{} in these scenarios. For TPC-C, \LEDS{} reduces the number of writes by 7.2$\times$ over \Manual{}.

\Auto{} reduces amplification and increases the write endurance of NVM by only updating the data that is accessed instead of copying the whole data structure.

\section{Related Work} 
\label{section:relatedwork}
To the best of our knowledge, there is no prior work that discusses a \FullName{} or a similar concept. 

There are several research projects focusing on programming interfaces for persistent memory, and here we discuss relevant works not previously discussed in the paper. Recent research like \cite{mnemosyne,nvheaps,Boehm:2016,nvmdirect} seeks a lightweight programming model to support programming with NVMM. Mnemosyne~\cite{mnemosyne} provides simple interfaces for the C language and using software transactional memory to provide atomicity. NVHeaps~\cite{nvheaps} provides persistent objects with transactional semantics while preventing possible pointer-related errors. In \cite{Boehm:2016}, they systematically explore a programming model suitable for persistent programming by defining semantics and identifying implementation costs for a wide range of programs. They do mention that persistent programming models need to adopt a model for checking if it can restart from the persistent data or not. 

There are more prior works, like \cite{Andrews:1987, Lamb:1991, Singhal:1992, Carey:1994, Atkinson:1996, Liskov:1996}, that provide persistent programming to database systems. We believe that while the implementation of our design is based on PMDK, our approach can be applied in a variety of existing persistent programming libraries that provide dynamic persistent memory allocation and atomicity primitives.

Another technique that bears resemblance to ours is Dynamic Software Updating (DSU)~\cite{hicks-dsu}. DSU, at its heart, is also about retaining data from one version of software to another. 
But our work has different goals. Their work focuses on upgrading a running program, like a server that cannot be brought down. However, \NameShort{} focuses on giving programmers a means of reasoning about outcomes when developing programs with persistent data. Another difference is that DSU requires programmers to provide both versions of the program and a dynamic patch in order for compilers to fix the memory during compilation, whereas \NameShort{} doesn't require full source code of previous versions and only needs the layout of the data that changes.

\LEDS{} bears similarity to structure splitting~\cite{hagog2005cache, golovanevsky2007struct}.

\section{Conclusion}
\label{section:conclusion}
We identify an important dimension introduced by persistent programs: how to retain persistent data in the presence of data layout changes. A key challenge is giving the programmer a way to reason about such changes. We refer to this as the \FullName{}, and we propose \LEDS{}, an example of the \Auto{}. We show that \LEDS{} offers competitive performance and significantly fewer writes than the state-of-the-art for the workloads studied.

\section*{Acknowledgements}
Wang and Tuck were supported in part by NSF grant CNS-1717486 and by NC State.

\bibliographystyle{plain}
\bibliography{ms}

\end{document}